\documentclass{article}
\usepackage{amsfonts,amssymb, amsmath}
\usepackage{float}
\usepackage{graphicx}
\usepackage{hyperref}

\begin{document}

\title{On rheonomic nonholonomic deformations of the Euler equations proposed by Bilimovich}
\author{A.V. Borisov, A.V. Tsiganov\\
Steklov Mathematical Institute, Russian Academy of Sciences\\
borisov@rcd.ru, andrey.tsiganov@gmail.com
}%
\date{}
\maketitle

\begin{abstract}
In 1913 A.D. Bilimovich observed that rheonomic   linear  and  homogeneous  in generalized velocities  constraints are ideal. As a typical example, he  considered rheonomic  nonholonomic deformation of the Euler equations which  scleronomic version is equivalent to the nonholonomic Suslov system. For the Bilimovich system equations of motion are reduced to quadrature, which is discussed in rheonomic and scleronomic cases.
\end{abstract}


\maketitle

\section{Introduction}
Let $q=(q_1\ldots,q_n)$  be  generalized  coordinates  on  the  configuration  space $Q$  of  the  system.
The Lagrange equations describing the motion of the system may be written as
\[
\dfrac{d}{dt}\,\dfrac{\partial T}{\partial \dot{q_i}}-\frac{\partial T}{\partial q_i}=Q_i\,,\qquad i=1,\ldots,n,
\]
where $T$ denotes the kinetic energy and $Q$ is a force.  Assume now that the  system is subject to additional independent constraints
\[f_j(q,\dot{q},t)==0,\qquad  j=1,\ldots,k <n\]
and we have a constrained Lagrangian system with the number of degrees of freedom dim$Q-k=n-k$. These
 constraints may be thought of as an addition of constraint forces to the original  Lagrange equations.

The constraint is called integrable if it can be written in the form
 \[
 f_j=\frac{d}{dt} g_j(q,t)=0
 \]
 for a convenient function $g_j$. Otherwise the constraint is called nonintegrable. After 1917, according to Hertz, nonintegrable constraints have been called nonholonomic.

Similarly to  Lagrangian function, constraints may be time-dependent  (rheonomic) or t\-ime\--\-in\-de\-pen\-dent (scleronomic). Thus, we say that the constrained Lagrangian system is scleronomic (rheonomic) if the constraints and the Lagrangian are time-independent (time-dependent).

A nonholonomic constraint $f_j=0$ is said to be ideal if the infinitesimal work of the constraint force vanishes for
any admissible infinitesimal virtual displacement
\begin{equation}\label{null-work}
\sum_{i=1}^k\frac{\partial f_j}{\partial\dot{q}_i}\delta{q}_i=0\,.
 \end{equation}
This equation is the so-called  Chetaev condition (see the paper \cite{chet32}, published in 1932). Some simple examples show that Chetaev's rule cannot be used in general (see \cite{cen06,mar98}).

Equations of motion of the nonholonomic system are deduced using  the Lag\-ran\-ge-d'Alembert principle,  Gauss and Appel principles,  Hamilton-Suslov principle and so on.
The general theory of linear and nonlinear, rheonomic and scleronomic, ideal and nonideal constraints and the corresponding nonholonomic systems is discussed in many recent
papers and textbooks. From the existing extensive list of literature we have chosen publications, particularly close to the work of Bilimovich \cite{bil,bil1},
see \cite{ang02,bmb16,dr07, fed09,fr14,dl99,kob04,nf72,ob10,pop14,ran94,rum06} and references
therein.

In 1903 A.D. Bilimovich  graduated from Kiev University with a first degree diploma and a gold medal for his work ''Application of geometric derivatives to the theory of curves and surfaces''. Here he becomes a student of famous mechanics K. G. Suslov and P.V. Voronets. After graduation, he was left as a fellowship and a freelance assistant to the Department of Theoretical and Applied Mechanics. In 1907 he received the title of Privatdocent of the Department of Theoretical and Applied Mechanics of Kiev University, where in 1912 he defended his master's thesis "Equations of motion for conservative systems and their applications"\,, which was published later in two papers \cite{bil,bil1}.

After the death of A. M. Lyapunov on November 3, 1918, he headed the commission for the preservation, processing and preparation for printing of the academician's works, which saved his manuscript ''On Some Equilibrium Figures of Rotating Fluid''.

In January 1920, he left Odessa and soon found refuge in Serbia, where he created a large scientific school in analytical mechanics. He also owes the merit of creating a number of scientific associations and institutes in Serbia (in the 1920-1960s), participation in publishing activities of compatriots-immigrants: the Russian Academic Circle (April), the Russian Scientific Institute, two editions of ''Materials for Russian Bibliography scientific works abroad '', Mathematical Institute of the Serbian Academy of Sciences, the opening of which took place in May 1946. In 1949, the first volume of  "Transactions of the Mathematical Institute of the Serbian Academy of Sciences ''\, was published. It was in this edition for several years that he published his works, including the memoirs of Lyapunov in Odessa (1956). In addition, A.D. Bilimovich was one of the founders of the Yugoslav Society of Mechanics. Scientific activity was marked by his election on February 18, 1925 as a corresponding member, and February 17,  1936 as a full member of the Serbian Academy of Sciences and Arts  \cite{bil2}.

So, in 1913-1914   Bilimovich published two papers in which discuss "commonly-known fact"\, that scleronomic constraints
 \[f_j=\sum_{i=1}^n b_{ij}(q)\dot{q}_j+b_j=0,\qquad  j=1,\ldots,k <n\,,
 \]
are  ideal if $b_j=0$.  Bilimovich noted  that rheonomic constraints linear  and  homogeneous  in generalized velocities
 \[f_j=\sum_{i=1}^n b_{ij}(q,t)\dot{q}_j=0,\qquad  j=1,\ldots,k <n\,,
 \]
 are also ideal constraints.  First  Bilimovich's paper \cite{bil} was submitted to Comptes rendus de l'Acad\'{e}mie des Sciences  by Appel, so Appel also knew this "commonly known fact" (\ref{null-work}).  Thus, Appel, Suslov and Voronets also know this "commonly-known fact" (\ref{null-work}) for linear  and  homogeneous  in generalized velocities  constraints but we do not find the original source of this fact.

  A typical example of  rheonomic constraints is a rod of varying length, see  Bilimovich's paper \cite{bil1}. In this note we discuss another typical example associated with a rotating rod, which allows us to study  rheonomic  nonholonomic deformations of the Euler equations \cite{bil}.  This rheonomic Bilimovich system is a generalization of the scleronomic nonholonomic Suslov system up to the choice of coordinates.  The physical realization of the corresponding constraint which was  proposed by Bilimovich does not realistic, in contrast to the nonholonomic pendulum \cite{bil1}, but  it is also a typical rheonomic nonholonomic system, which we will integrate by quadratures below.

\subsection{On constraints that are linear in velocities and imposed on a conservative dynamical system}
Consider Lagrange's equations of the second kind
\begin{equation}\label{lag-eq}
\dfrac{d}{dt}\,\dfrac{\partial T}{\partial \dot{q_i}}-\frac{\partial T}{\partial q_i}=Q_i\,,\qquad i=1,\ldots,n,
\end{equation}
in which the kinetic energy is a second-degree polynomial in the generalized velocities and can be represented as
\[
T=T_2+T_1+T_0\,,
\]
where
\[
T_2=\frac12\sum_{i,j=1}^{n} a_{ij}\dot{q}_i\dot{q_j}\,,\qquad T_1=\sum_{k=1}^{n}a_k\dot{q}_k\,,\qquad T_0=a_0\,,
\]
and the coefficients $a_{ij}$ and $a_k$ are functions of the generalized coordinates $q_1,\ldots,q_n$ and time $t$.

If one divides the forces $Q_i$ into potential and nonpotential ones
\[
Q_i=-\frac{\partial V}{\partial q_i}+\widetilde{Q}_i\,,
\]
then one can formulate a theorem of the change in the total mechanical energy $H=T+V$ of the holonomic system
\begin{equation}\label{teor-H}
\dfrac{d }{dt} H=\sum_{i=1}^n \widetilde{Q}_i\dot{q}_i+\frac{d}{dt} (T_1+2T_0)-\frac{\partial H }{\partial t}\,.
\end{equation}
The imposition of $k<n$ nonholonomic constraints
\[f_j( q,\dot{q})=0,\qquad j=1,\cdots, k<n\]
changes Lagrange's equations (\ref{lag-eq})
\begin{equation}\label{lag-eq-g}
\dfrac{d}{dt}\,\dfrac{\partial T}{\partial \dot{q_i}}-\frac{\partial T}{\partial q_i}=Q_i+\sum_{j=1}^k \lambda_j\frac{\partial f_j}{\partial \dot{q}_i}\,,\qquad i=1,\ldots,n,
\end{equation}
and the theorem of the change in the total mechanical energy(\ref{teor-H})
\begin{equation}\label{teor-H-g}
\dfrac{d}{dt} H=\sum_{i=1}^n( \widetilde{Q}_i+Q_i^*)\dot{q}_i+\frac{d}{dt} (T_1+2T_0)-\frac{\partial H }{\partial t}\,,\qquad
Q_i^*=\sum_{j=1}^k\lambda_j\sum_{i=1}^n \frac{\partial f_j}{\partial \dot{q}_i}\dot{q}_i\,,
\end{equation}
where $Q_i^*$ are the reaction forces of the nonholonomic constraints.

The homogeneous linear functions of the generalized velocities
\[
f_j=\sum_{i=1}^n b_{ji}(q,t) \dot{q}_i\,
\]
are distinguished by the fact that in this case the imposition of nonholonomic constraints does not change the theorem of the change in the total mechanical energy, since
the work done by these corresponding reaction forces is zero
\[
\sum_{i=1}^n Q_i^*\dot{q}_i=\sum_{j=1}^k\lambda_j\sum_{i=1}^n \frac{\partial f_j}{\partial \dot{q}_i}\dot{q}_i=
\sum_{j=1}^k\lambda_j\sum_{i=1}^n b_{ji}\dot{q}_i=\sum_{j=1}^k\lambda_jf_j=0\,.
\]
This fact follows from Euler's theorem of homogeneous functions, which is also used for a standard derivation of the initial theorem of the change in the total mechanical energy of the
holonomic system.

In discussing this well-known fact, in \cite{bil}, Bilimovich noted that the coefficients $b_ {ij} $ can explicitly depend on time $t$. Thus, imposing
rheonomic constraints linear in the velocities
on the
conservative system whose total mechanical energy does not change during the motion of the system
\[
\dfrac{d}{dt}H=0\qquad\Rightarrow\qquad H=E\,,
\]
we obtain a rheonomic nonholonomic system for which the total mechanical energy is conserved as well: $H=E$.

\section{Rheonomic deformations of the Euler equations}
The main example in the work of Bilimovich \cite{bil} is related to the nonholonomic deformation of Euler's equations
\begin{equation}\label{eul-eq}
I\dot{\omega}=I\omega \times \omega\qquad \Longleftrightarrow \qquad
\frac{d}{dt}\left(\frac{\partial T}{\partial \omega_i}\right)=Q_i=(I\omega \times \omega)_i\,,
 \end{equation}
where $\omega$ is the angular velocity vector of a rigid body and $I$ is the inertia tensor of the body. For this model conservative system the total
mechanical energy coincides with the kinetic energy
\begin{equation}\label{ham-bil}
T=\frac12(I\omega,\omega)\,,
\end{equation}
which remains unchanged during motion, as does the squared angular momentum of the body, $M^2$.

Solving two equations $T=E$ and  $M^2=m$ for $\omega_1$ and $\omega_2$ and substituting the resulting solutions into the equation of motion for the third angular velocity
component, we obtain the well-known autonomous differential equation with separable variables
\[
\dfrac{d \omega_3}{dt}=\sqrt{F(\omega_3,E,m)} or \left(\dfrac{d \omega_3}{dt}\right)^2=F(\omega_3,E,m)\,,
\]
which contains a fourth-degree polynomial $F$ in $\omega_3$. An explicit form of this polynomial and a solution to this quadrature in terms of elliptic functions can be found,
for example, in the textbooks \cite{bm01,fr14}.

Imposing on this integrable system the nonholonomic rheonomic constraint
\begin{equation}\label{f-con}
f=\omega_1-g(t)\,\omega_2=0\,,
\end{equation}
where $g(t)$ is an arbitrary function of time, changes the equations of motion
 \begin{equation}\label{bil-eq}
I\dot{\omega}=I\omega\times \omega+\lambda \dfrac{\partial f}{\partial \omega}\,,
\qquad \Longleftrightarrow \qquad
\frac{d}{dt}\left(\frac{\partial T}{\partial \omega_i}\right)=(I\omega \times \omega)_i+\lambda \dfrac{\partial f}{\partial \omega_i}\,,
 \end{equation}
 in which, however, the total mechanical energy of the system remains
unchanged:
\[
T=\frac12(I\omega,\omega)=E\,.
\]
The undetermined Lagrange multiplier $\lambda$ appearing in these equations can be found by differentiating the constraint
\begin{equation}\label{lag-mul1}
\frac{df}{dt}=\sum_{i=1}^3 \frac{\partial f}{\partial \omega_i}\dot{w}_i+\frac{\partial f}{\partial t}=0\,.
\end{equation}
 Solving the equations $T=E$ and $f=0$ for $\omega_1$ and $\omega_2$, we obtain
 \[
 \omega_1 =\dfrac{ \sqrt{A}-\bigl(g(t)I_{13}+I_{23}\bigr)\omega_3}{C}\,g(t)\,,\qquad
 \omega_2 =\dfrac{ \sqrt{A} -\bigl(g(t)I_{13}+I_{23}\bigr)\omega_3}{C}\,,
  \]
 where
 \[\begin{array}{rcl}
 A&=&2EC -B\omega_3^2\,,\qquad C=g(t)^2 I_{11} + 2g(t) I_{12} + I_{22}\,,
 \\
 \\
 B&=&g(t)^2 (I_{11}I_{33} - I_{13}^2) +2g(t) (I_{12}I_{33} - I_{13}I_{23}) + I_{22}I_{3 3} - I_{23}^2\,.
 \end{array}
 \]
Substituting these solutions into the equation of motion for $\omega_3$, we obtain a nonautonomous differential equation with separable variables
\begin{equation}\label{gen-eq}
\begin{array}{rcl}
\frac{d \omega_3}{dt}&=&\frac{\bigl((1-g^2)I_{12} + g(I_{11} -I_{22})\bigr)A}
{BC}
\\
\\
&-&\frac{
\Bigl(\bigl((1+g^2)\omega_3 - g'\bigr)\sqrt{A}  -g'\omega_3(gI_{13} +I_{23}) \Bigr)  \bigr(g(I_{11}I_{23} -I_{12}I_{13}) + I_{12}I_{23} - I_{13}I_{22}\bigl)
}
{BC}\,,
\end{array}
\end{equation}
where $g=g(t)$ and $g'=dg(t)/dt$.  Similar to the Euler case it can be rewritten in a square-free form
\begin{equation}\label{gen-eq-f}
\begin{array}{rcl}
\left(
\frac{d \omega_3}{dt}-\frac{\bigl((1-g^2)I_{12} + g(I_{11} -I_{22})\bigr)A}
{BC}\right.
&-&\left.\frac{
\Bigl(g'\omega_3(gI_{13} +I_{23}) \Bigr)  \bigr(g(I_{11}I_{23} -I_{12}I_{13}) + I_{12}I_{23} - I_{13}I_{22}\bigl)}
{BC}\right)^2
\\
\\
&=&\frac{\bigl((1+g^2)\omega_3 - g'\bigr)^2 \bigr(g(I_{11}I_{23} -I_{12}I_{13}) + I_{12}I_{23} - I_{13}I_{22}\bigl)^2A}{BC}\,.
 \end{array}
\end{equation}
The similar equation for rheonomic nonholonomic pendulum Bilimovith reduced  to elliptic quadrature after suitable change  of time \cite{bil1}. We suppose that this equation (\ref{gen-eq-f}) could be also reduced to  to elliptic quadrature. If one integrates this equation, then the remaining components of the angular velocity vector, $\omega_1(t)$ and $\omega_2(t)$, can be found from the equations $T=E$ and $f=0$.

Thus, we obtain formal quadrature (\ref{gen-eq-f}) for the abstract rheonomic Lagrangian system with constraint (\ref{f-con}). Of course, in generic we have to define more exactly physical properties of the constraints, which impose restrictions to the set of possible values of the constraint forces and to the admissible path. In order to do it, we need  a realistic physical realization of the constraints, which is absent for the  Bilimovich system.

In  \cite{bil}  Bilimovich studied  tensor of inertia
 \[
 I=\left(
    \begin{array}{ccc}
      \mathrm A & 0 & 0 \\
      0 & \mathrm A & 0 \\
      0 & 0 & \mathrm C \\
    \end{array}
  \right)\,,
 \]
 and proved that integration of the equations of motion (\ref{bil-eq}) is reduced to the simultaneous integration of the following equations in term of the Euler angles
\[\begin{array}{c}
\dot{\psi}\sin\phi \left[-\cos\theta+g(t)\sin\theta\right]+\dot{\phi}\left[\sin\theta+g(t)\cos\theta\right]=0\,,
\\
\\
\dot{\psi}\cos\phi+\dot{\theta}=\Gamma\,,
\\
\\
\mathrm A\left(\dot{\psi}^2\sin^2\phi+{\dot{\phi}}^2\right)=2E-\mathrm C\Gamma^2\,,
\end{array}
\]
where $\Gamma$ is an arbitrary constant. Of course, we can also rewrite our quadrature (\ref{gen-eq}) in term of the Euler angles using  change of variables
\[\begin{array}{c}
\omega_1=p=-\dot{\psi}\sin\phi\cos\theta+\dot{\phi}\sin\theta\,,\\ \\
\omega_2=q=\dot{\psi}\sin\phi\sin\theta+\dot{\phi}\cos\theta\,,\\ \\
\omega_3=r\dot{\psi}\cos\phi+\dot{\theta}
\end{array}
\]
from the  Bilimovich paper \cite{bil}.

Let us briefly discuss solutions of  (\ref{gen-eq}) in some partial cases.
For example, if the tensor of inertia has the form
\[
I=\left(
    \begin{array}{ccc}
      I_{11} & I_{12} & 0 \\
      I_{12} & I_{22} & 0 \\
      0 & 0 & I_{33} \\
    \end{array}
  \right)\,,
\]
then the second term in ({\ref{gen-eq}) disappears and separated equation becomes a square root free
\[
\frac{d\omega_3}{dt}=
\frac{(2E-I_{33}\omega_3^2)\Bigl(I_{12}(1-g(t)^2) +\bigl(I_{11}-I_{22}\bigr)g(t)\Bigr)}{I_{33}(I_{11} g(t)^2 + 2I_{12}g(t) + I_{22})}.
\]
The general solution to this equation is
\[
\omega_3(t)=\sqrt{\frac{2E}{I_{33}}}\tan\left(\sqrt{\frac{2E}{I_{33}}}\left(c-
\int\frac{\Bigl(I_{12}(1-g(t)^2) +\bigl(I_{11}-I_{22}\bigr)g(t)\Bigr)}{I_{11} g(t)^2 + 2I_{12}g(t) + I_{22}}dt
\right)\right)\,,
\]
where $c$ is the constant of integration.

For example, at $ E = 1, I_{33} = 4, I_{11} = 2, I_{22} = 1$ and $ I_{12}=I_{13}=I_{23} = 0$, solutions to the separated equation (\ref{gen-eq}) with  $g(t)=\cos(t)$ and
$g(t)=\alpha$ are presented in Fig.1.
\begin{figure}[!ht]
\center{\includegraphics[width=0.7\linewidth, height=0.3\textheight]{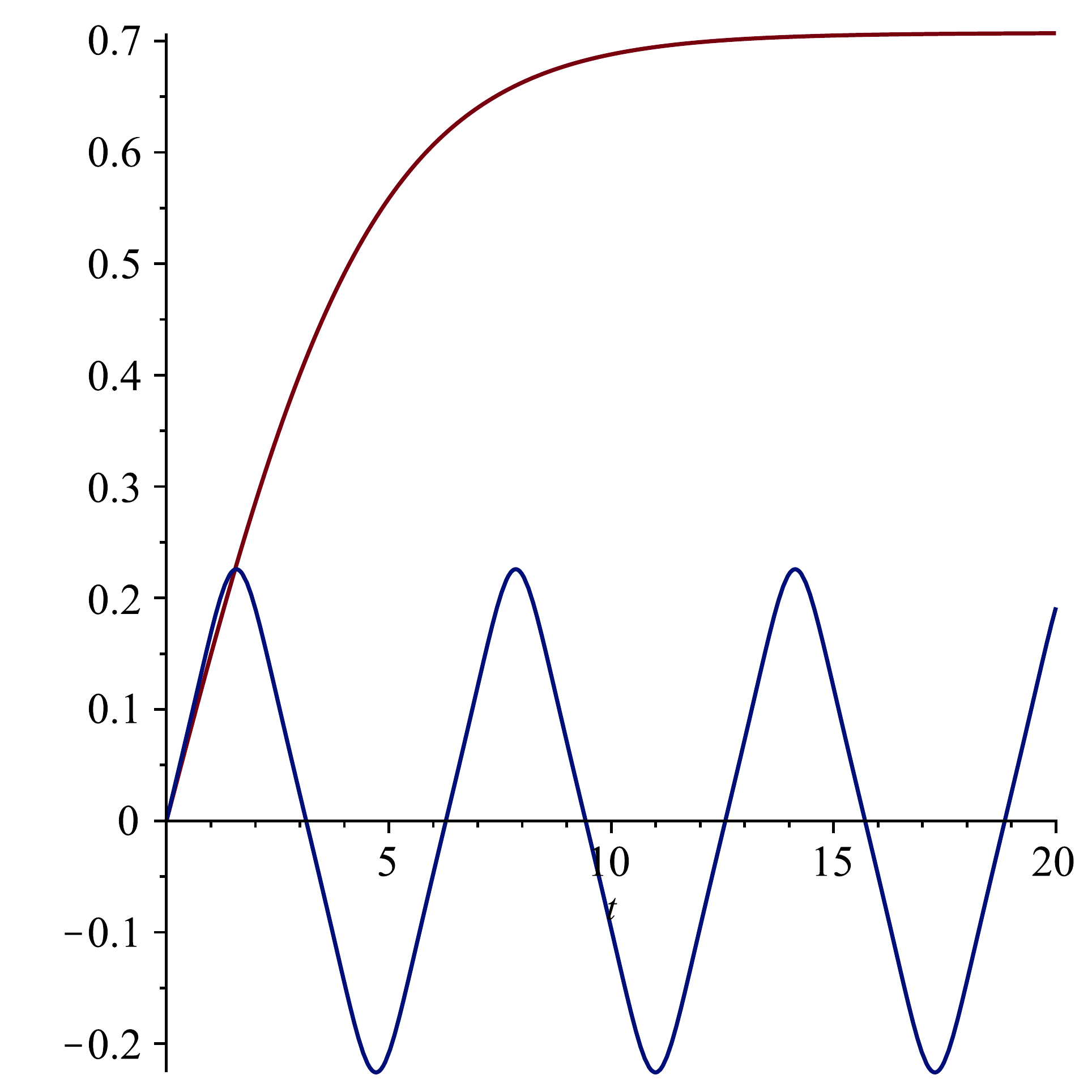} }
\caption{Graphs of $\omega_3(t)$ at $ I_{12}=I_{13}=I_{23} = 0$. }
\end{figure}
\par\noindent
The red curve denotes a graph for the scleronomic constraint $g(t)=0.4$, and the blue curve is a solution graph for the rheonomic constraint  with the periodic function $g(t)=\cos t$.

If at the same parameter values the off-diagonal moment of inertia $ I_{12}$ does not zero, for example, $ I_{12}=0.05$, then the solutions to the separated equation (\ref{gen-eq})
with $g(t)=\cos(t)$ and $g(t)=\alpha$ have the
following form:
\begin{figure}[H]
\center{\includegraphics[width=0.7\linewidth, height=0.3\textheight]{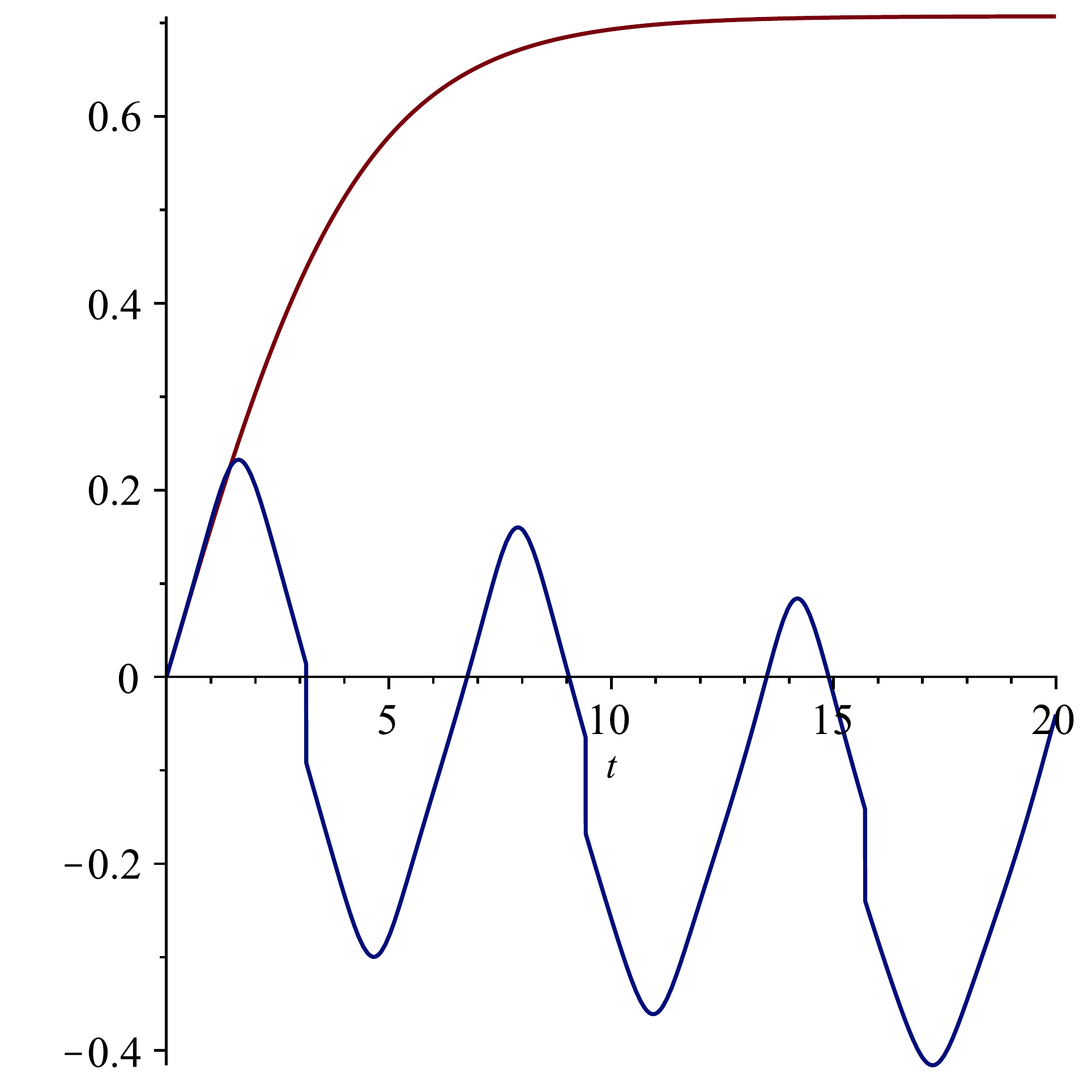} }
\caption{Graphs of $\omega_3(t)$ at $ I_{12}=0.05$, $I_{13}=I_{23} = 0$.}
\end{figure}
For rheonomic constraint the corresponding phase portrait looks like
\begin{figure}[H]
\center{\includegraphics[width=0.7\linewidth, height=0.3\textheight]{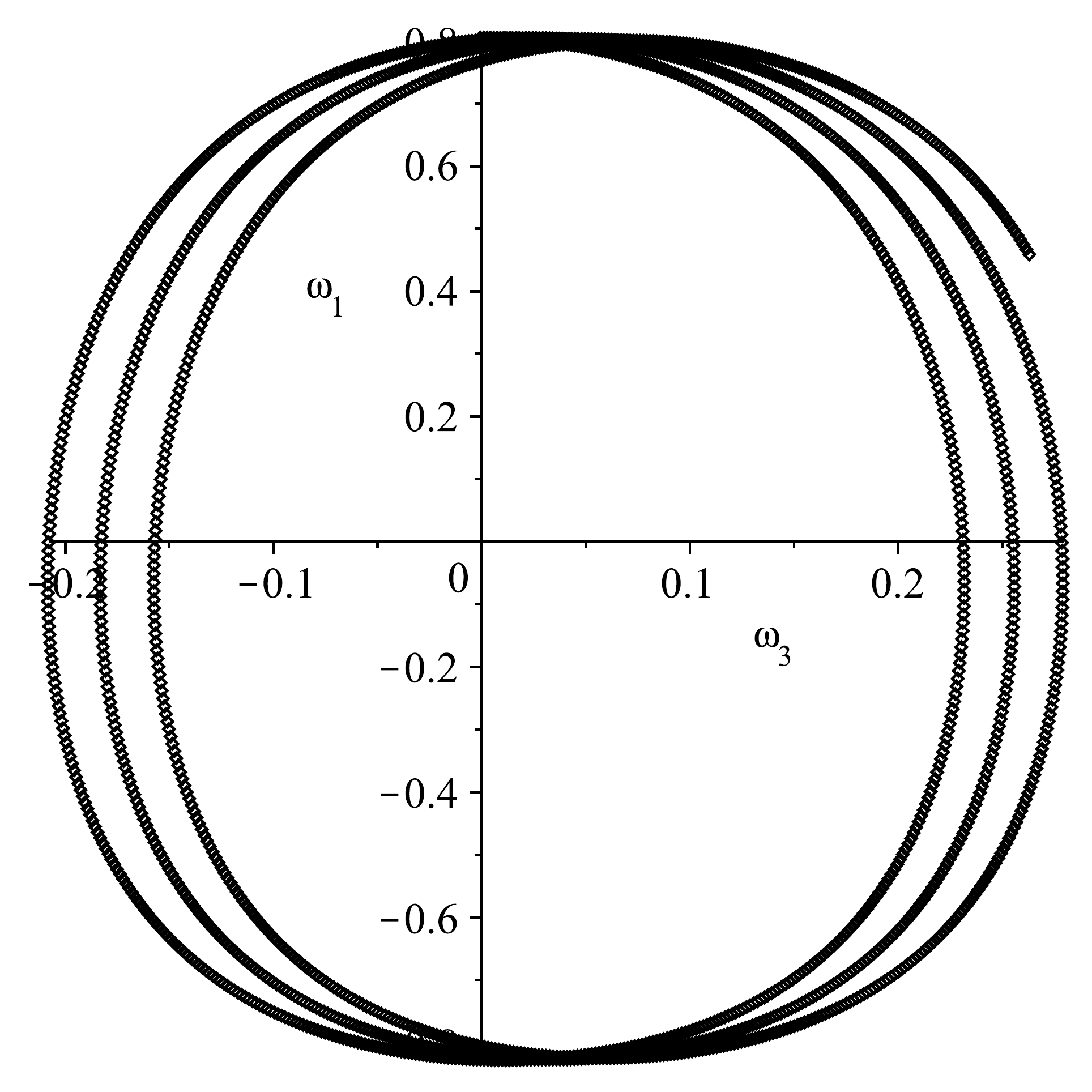} }
\caption{Dependence $\omega_1$ on $\omega_3$ at $ I_{12}=0.05$, $I_{13}=I_{23} = 0$.}
\end{figure}
\par\noindent
If $I_{13}\neq 0$ and $I_{23}\neq0$, the solutions to the separated equations (\ref{gen-eq}) or (\ref{gen-eq-f}) can be obtained numerically. For instance, if $ I_{12}=0.05$ and $I_{13}=-1$, then the numerical solutions to the separated equation (\ref{gen-eq}) with
$g(t)=\cos(t)$ and $g(t)=\alpha$ have the following form at $A>0$
\begin{figure}[H]
\center{\includegraphics[width=0.7\linewidth, height=0.3\textheight]{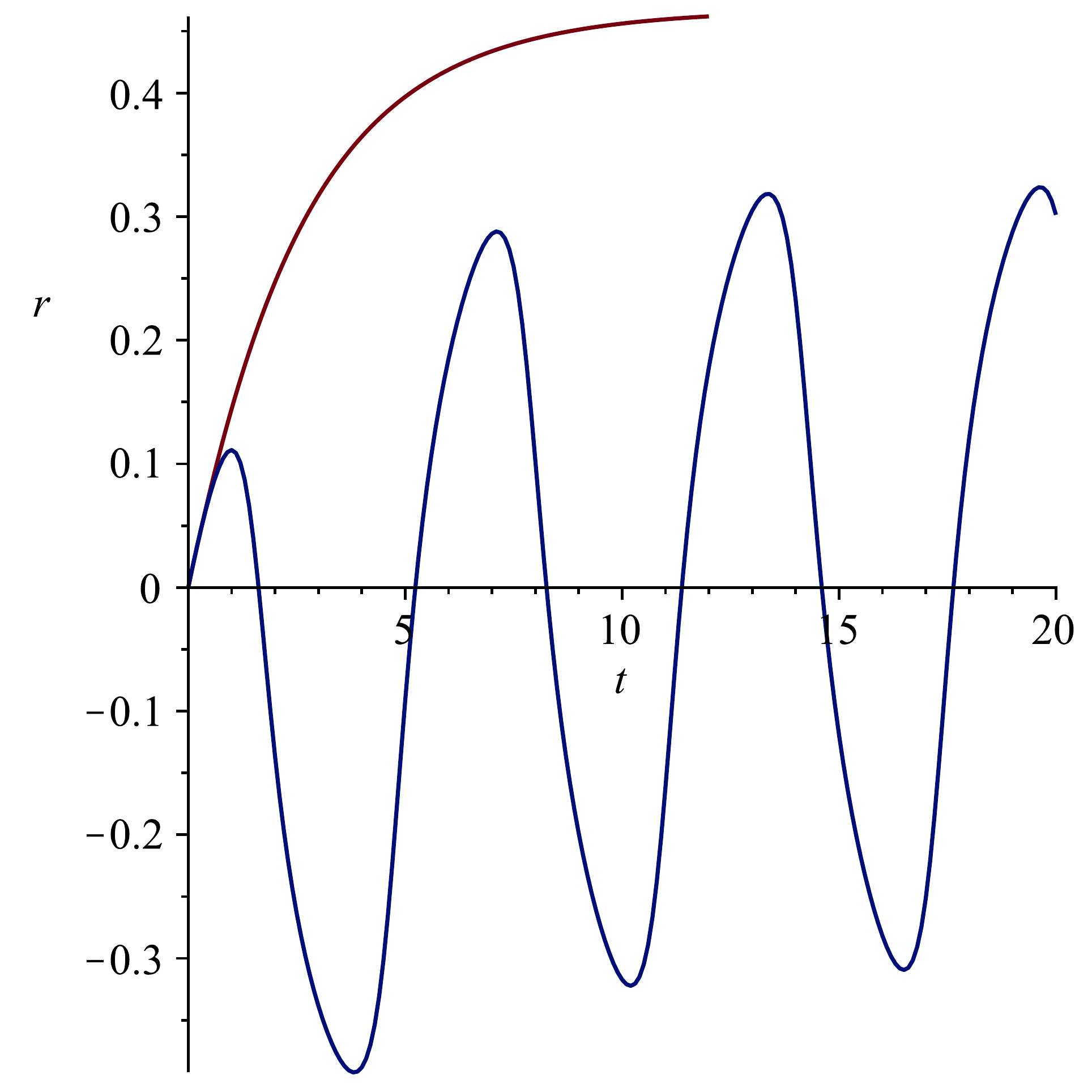} }
\caption{Graphs of $\omega_3(t)$ at $ I_{12}=0.05$, $I_{13}=-1$ and $I_{23} = 0$.}
\end{figure}
For rheonomic constraint the corresponding phase portrait looks like
\begin{figure}[H]
\center{\includegraphics[width=0.7\linewidth, height=0.3\textheight]{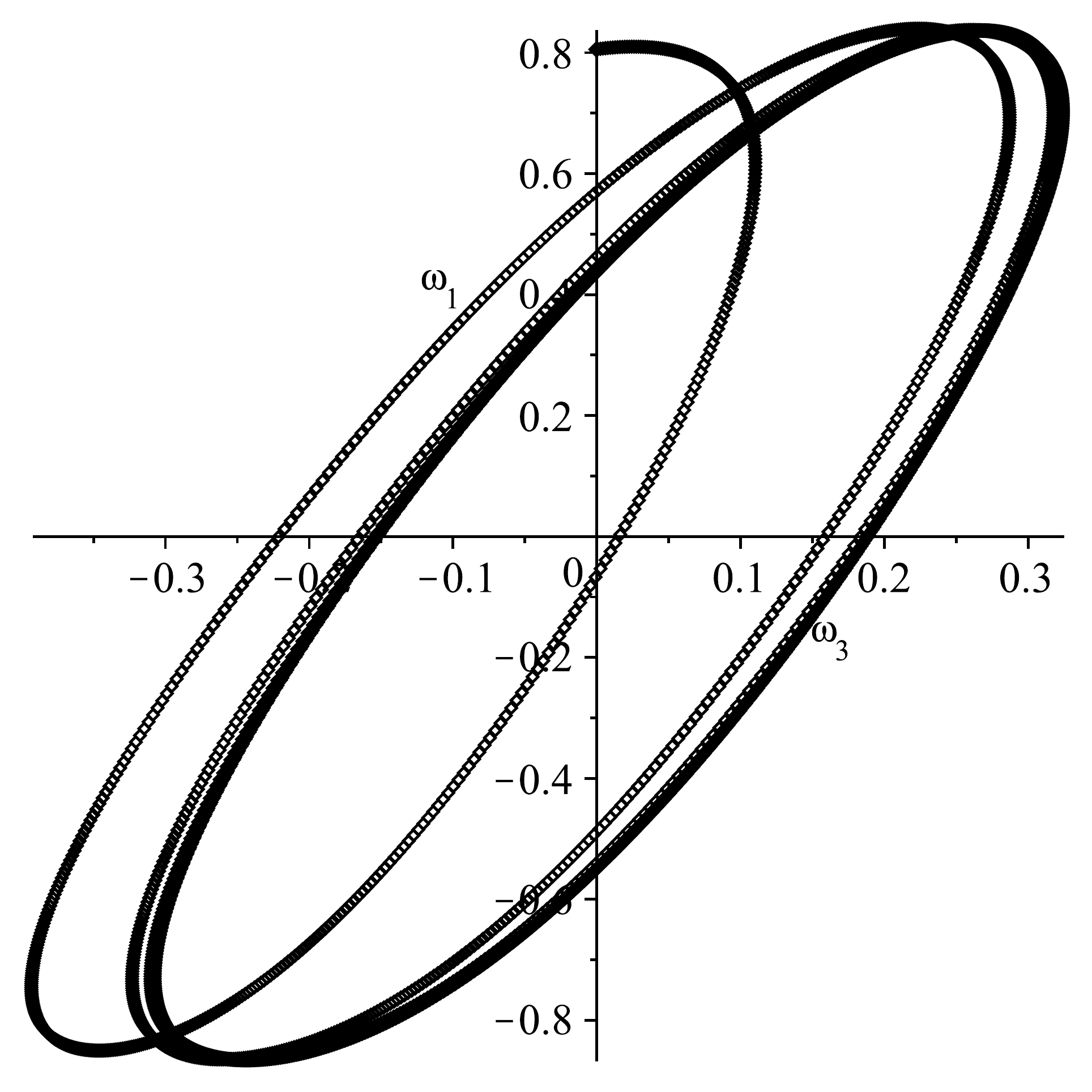} }
\caption{Dependence $\omega_1$ on $\omega_3$ at $ I_{12}=0.05$, $I_{13}=-1$ and $I_{23} = 0$.}
\end{figure}
\par\noindent
After some time the motion becomes almost periodic. Here $I_{13}\neq = 0$, but equation (\ref{gen-eq}) can be integrated  numerically because $A>0$ under selected initial conditions. At the other values of parameters and initial conditions we have to solve second order nonlinear equation (\ref{gen-eq-f}) which can not be converted to an explicit first-order system.

The off-diagonal moments of inertia do almost not change the motion pattern in the case of the scleronomic constraint.
In the case of the rheonomic constraint, the motion pattern depends on the values of the off-diagonal moments of inertia
rather strongly.  In what follows, we will show how in the scleronomic case the off-diagonal moments of inertia are related to the existence of an invariant measure and to the Hamiltonian property of equations of motion.

 \subsection{Scleronomic deformations of the Euler equations}
If we assume that the function $g(t)$ is constant, $g(t)=\alpha$, then the system of three differential equations
(\ref{bil-eq})
\[
X=
\left(
\begin{array}{c}
   \dot{\omega}_1 \\
    \dot{\omega}_2 \\
    \dot{\omega}_3 \\
  \end{array}
\right)
= I^{-1}\left(
I\omega\times \omega\right)+\lambda I^{-1}\left(
                               \begin{array}{c}
                                 1 \\
                                 -\alpha \\
                                 0 \\
                               \end{array}
                             \right)
\]
possesses two integrals $T$ and $\alpha=\omega_1/\omega_2$. This makes it possible to reduce the order of this system
by two and to obtain one differential equation (\ref{gen-eq}), which in this case has the form
\begin{equation}\label{a-F}
\begin{array}{rcl}
\frac{d\omega_3}{dt}&=&F(\omega_3,E,\alpha)
\\
\\
&=&\frac{(\alpha^2I_{12} -\alpha(I_{11}-I_{22}) -I_{12})A}{BC}+\frac{(\alpha^2 + 1)
\bigl(\alpha(I_{11}I_{23} -I_{12}I_{13}) + I_{12}I_{23} -I_{13}I_{22}\bigr)\omega_3\sqrt{A\,}}{BC},
\end{array}
\end{equation}
where
\[\begin{array}{rcl}
A&=&B \omega_3^2 + 2EC\,,\qquad C=(\alpha^2I_{11} + 2\alpha I_{12} + I_{22})\,,
\\
\\
B&=&\alpha^2( I_{13}^2-I_{11}I_{33} -) + 2\alpha(I_{13}I_{23}-I_{12}I_{33} ) - I_{22}I_{33}+ I_{23}^2.
\end{array}
\]
Note that calculations simplify considerably at
\begin{equation}\label{eq-mera}
\alpha(I_{11}I_{23} -I_{12}I_{13}) + I_{12}I_{23} -I_{13}I_{22}=0
\end{equation}
when in the definition of the function $F$ (\ref{a-F}) the root $\sqrt{A}$ is absent and
\[
F=\frac{(\alpha^2I_{12} -\alpha(I_{11}-I_{22}) -I_{12})A}{BC}\,.
\]
Under the condition det$I\neq 0$ this simplification occurs in the unique case $I_{13}=I_{23}=0$, which we will
consider below in more detail.

When $I_{13}=I_{23}=0$, the Lagrange multiplier is
\[\scriptstyle
 \lambda = -\frac{\omega_2\omega_3\Bigl(
 (I_{11}^2 - I_{11}I_{33} + I_{12}^2)\omega_1^2 + 2\omega_1\omega_2I_{12}(I_{11} + I_{2 2} - I_{3 3})\omega_2 + (I_{1 2}^2 + I_{2 2}^2 - I_{2 2}I_{3 3})\omega_2^2
 \Bigr)}
 {I_{11}\omega_1^2+2I_{12} \omega_1 \omega_2 + I_{22}\omega_2^2}\,.
 \]
Substituting $\lambda$ into the equations of motion  (\ref{bil-eq}), we obtain the vector field
\[X=\left(
      \begin{array}{c}
        \dfrac{\bigl(\omega_1^2 I_{12} - \omega_2 (I_{11} - I_{22}) \omega_1 - \omega_2^2 I_{12}\bigr) \omega_1 \omega_3}{\omega_1^2 I_{11} + 2 \omega_1 \omega_2 I_{12} + \omega_2^2 I_{22}}
        \\ \\
        \dfrac{\bigl(\omega_1^2 I_{12} - \omega_2 (I_{11} - I_{22}) \omega_1 - \omega_2^2 I_{12}\bigr) \omega_2 \omega_3}{\omega_1^2 I_{11} + 2 \omega_1 \omega_2 I_{12} + \omega_2^2 I_{22}}
        \\ \\
        \dfrac{-\omega_1^2 I_{12} + \omega_2 (I_{11} - I_{22}) \omega_1 + \omega_2^2 I_{12}}{I_{33}} \\
      \end{array}
    \right)
\]
which preserves the Hamiltonian $T$ (\ref{ham-bil}) and the ratio between the velocities $\alpha=\omega_1/\omega_2$
and possesses the invariant multiplier
\[
\rho=\frac{1}{\omega_1\omega_2}\,,\qquad \sum_{i=1}^3 \frac{\partial \bigl( \rho X_i\bigr)}{\partial \omega_i}=0\,.
\]
which defines the invariant singular measure
\[
\mu=\rho(\omega_1,\omega_2,\omega_3)d\omega_1d\omega_2d\omega_3\,.
\]
In addition, the vector field $X$ is Hamiltonian
\[
X=PdT
\]
with respect to the Hamiltonian $T$ and the Poisson bivector
\[\begin{array}{rcl}
P&=&\omega_1\omega_2\left(
    \begin{array}{ccc}
      0 & I_3\omega_3 & -I_{12}\omega_1-I_2\omega_2 \\
      -I_3\omega_3 & 0 & I_{12}\omega_1+I_1\omega_1 \\
     I_{12}\omega_1+I_2\omega_2 &- I_{12}\omega_1-I_1\omega_1 & 0 \\
    \end{array}
  \right)
  \\
  \\
  &-&\dfrac{\omega_1^2 I_{12} - \omega_1 \omega_2 I_{11} + \omega_1 \omega_2 I_{22} - \omega_2^2 I_{12}}{(\omega_1^2 I_{11} + 2 \omega_1 \omega_2 I_{12} + \omega_2^2 I_{22}) I_{33}}
  \left(
    \begin{array}{ccc}
      0 & 0 & \omega_1 \\
      0 & 0 & \omega_2 \\
      -\omega_1 & -\omega_2 & 0 \\
    \end{array}
  \right)\,.
  \end{array}
\]
For nonphysical solutions to equation (\ref{eq-mera}), for example,
\[
I=\left(
    \begin{array}{ccc}
      I_{11} & 0 & I_{13} \\
      0 & 0 & 0 \\
      I_{13} & 0 &I_{33} \\
    \end{array}
  \right),
\]
an invariant measure exists also, and the equations of motion are Hamiltonian.

\section{Conclusion}
In \cite{bil1} Bilimovich found solutions of the rheonomic constrained Lagrangian system in term of elliptic quadratures. It is a first example of explicit integration of equations of motion of nonholonomic systems with rheonomic constraints.

In \cite{bil} Bilimovich studied  rheonomic nonholonomic deformations of the Euler equations and very briefly discussed explicit integration of this system  in the partial axially symmetric case. In this note we present  explicit integration of these equations of motion in generic cases.

It will be interesting to study inhomogeneous Bilimovitc systems with constraint
\[
f=\omega_1-g(t)\,\omega_2=a\,,\qquad a\in \mathbb R\,.
\]
It is non ideal constraint and the corresponding constrained equations of motion do not preserve total energy $T$. However, the corresponding two-dimensional flow could preserve some integral in partial cases  similar to the inhomogeneous Suslov problem \cite{nar14}.

\subsection*{Acknowledgments} {This work was supported by the Russian Science Foundation (project no.~19-71-30012) and performed at the Steklov Mathematical Institute of the Russian Academy of Sciences. The authors declare that they have no conflicts of interest.}


\begin{thebibliography}{10}

\bibitem{ang02}
 J. Angeles,
\newblock{\em Fundamentals of Robotic Mechanical Systems: Theory, Methods, and Algorithms}, Springer Science \& Business Media, 2002.

\bibitem{bil} A. D. Bilimovich
\newblock{\em Sur les syst\`{e}mes conservatifs, non holonomes
avec des liaisons d\'{e}pendantes du temps},  Comptes Rendus Acad. Sci. Paris,  v.156, pp.12-18,  1913,\\
available from \url{https://gallica.bnf.fr/ark:/12148/bpt6k3109m/f1218.image.langFR}

\bibitem{bil1}  A. D. Bilimovich
\newblock{\em  La pendule nonholonome}, Mat. Sb., 1914, vol. 29, no. 2, pp. 234-240.
available from \url{http://www.mathnet.ru/links/c15779b7b313cb1249d626cb3380a92e/sm6518.pdf}

\bibitem{bil2}
 Anton Dimitrija Bilimovi\v{c} \url{http://www.mi.sanu.ac.rs/History/bilimovic.htm};\\
 \url{https://www.wikidata.org/wiki/Q4086698}


\bibitem{bm01}
A.\,V. Borisov and  I.\,S. Mamaev,
\newblock{\em Rigid body dynamics,} De Gruyter Stud. Math. Phys., vol. 52, Berlin: De Gruyter, 2018. 

\bibitem{bmb16}
 A.V. Borisov, I.S. Mamaev, I.A. Bizyaev,	
\newblock{\em Historical and critical review of the development of nonholonomic mechanics: the classical period},
Regular and Chaotic Dynamics, 2016, v. 21, no. 4, pp. 455-476.

\bibitem{cen06}
H. Cendra, S. Grillo,
\newblock{\em  Generalized nonholonomic mechanics, servomechanisms and related brackets},
J. Math. Phys. v. 47 (2006), no. 2, 022902, 29 pp.

\bibitem{chet32}
 N.G. Chetaev,
 \newblock{\em  On Gauss principle},  Izv. Fiz.-Mat. Obshch. Kazan Univ. (3) , v.6 (1932-1933) pp. 68-71.

\bibitem{dr07}
D. Djuri\'{c},
\newblock{\em  On stability of stationary motion of a nonconservative nonholonomic rheonomic system},
European Journal of Mechanics - A/Solids, v.26, issue 6, pp.1029-1039, 2007.

\bibitem{fr14}
R. Featherstone,
\newblock{\em  Rigid body dynamics algorithms}, Springer, 2014.

\bibitem{fed09}
Yu.N. Fedorov,  A.J. Maciejewski, M.  Przybylska,
\newblock{\em The Poisson equations in the nonholonomic Suslov problem: integrability, meromorphic and hypergeometric solutions},
Nonlinearity, 2009, vol. 22, pp. 2231-2259, 2009.

\bibitem{nar14}
L.C. Garc\'{\i}a-Naranjo, A.J. Maciejewski, J.C. Marrero, M. Przybylska,
\newblock{\em The inhomogeneous Suslov problem}, Physics Letters A,
v.378, pp. 2389-2394, 2014.

\bibitem{kob04}
M. H. Kobayashi and W. M. Oliva,
 \newblock{em A note on the conservation of energy and volume in the setting of nonholonomic mechanical systems}, Qualitative Theory of Dynamical Systems, v.4, pp.383-411, 2004.

\bibitem{dl99}
M. De Le\'{o}n, J.C. Marrero, D.M. De Diego,
 \newblock{\em Time-dependent mechanical systems with non-linear constraints}. In: Szenthe J. (eds) New Developments in Differential Geometry, Budapest 1996. Springer, Dordrecht, 1999.

\bibitem{mar98}
C.-M. Marle,
\newblock{\em Various approaches to nonholonomic systems}, Rep. Math. Phys., v.42, pp. 211-29, 1998.

\bibitem{nf72}
J. Neimark, N. Fufaev,
\newblock{\em  Dynamics of nonholonomic systems}, Transactions of Mathematical Monographs,
v. 33, AMS, Providence, RJ, 1972.

\bibitem{ob10}
A. Obradovi\'{c}, V. \v{C}ovi\'{c}, M. Veskovi\'{c}, M. Dra\v{z}i\'{c},
\newblock{\em Brachistochronic motion of a nonholonomic rheonomic mechanical system}, Acta Mech.,
v.214, pp. 291-304, 2010.

\bibitem{pop14}
P. Popescu, C. Ida,
\newblock{\em Nonlinear constraints in nonholonomic mechanics},
 Journal of Geometric Mechanics, v.6, pp. 527-547, 2014.

\bibitem{ran94}
M.F. Ra\~{n}ada,
\newblock{\em Time-dependent Lagrangians systems: A geometric approach to the theory of systems with constraints},
 J. Math. Phys., v.35, pp.748-758, 1994.

\bibitem{rum06}
V. Rumyantsev,
\newblock{\em  Variational principles for systems with unilateral constraints},
 Journal of Applied Mathematics and Mechanics,  v.70, pp.808 --818, 2006.

\end{thebibliography}
\end{document}